\documentclass[
reprint,
amsmath,
amssymb,
aps,
prapplied,
]{revtex4-2}

\usepackage{graphicx}
\usepackage{dcolumn}
\usepackage{bm}
\usepackage[colorlinks=true, allcolors=black]{hyperref}
\usepackage{physics}
\usepackage{nameref}
\usepackage{upgreek}

\begin{document}

\title{Optimizing Pulse Shapes of an Echoed Conditional Displacement Gate in a Superconducting Bosonic System}

\author{%
	Maxime Lapointe-Major,\textsuperscript{1,2,3} Yongchao Tang,\textsuperscript{1} Mehmet Canturk,\textsuperscript{1} and Pooya Ronagh\textsuperscript{1,4,5,6,}
}
\thanks{Corresponding author: \href{mailto:pooya.ronagh@1qbit.com}{pooya.ronagh@1qbit.com}}

\affiliation{\textsuperscript{\emph{1}}1QB Information Technologies (1QBit), Vancouver, BC, Canada\\
\textsuperscript{\emph{2}}D\'{e}partement de g\'{e}nie \'{e}lectrique et de g\'{e}nie informatique, Universit\'{e} de Sherbrooke, Sherbrooke, QC, Canada\\
\textsuperscript{\emph{3}}Institut Quantique, Universit\'{e} de Sherbrooke, Sherbrooke, QC, Canada\\
\textsuperscript{\emph{4}}Institute for Quantum Computing, University of Waterloo, Waterloo, ON, Canada\\
\textsuperscript{\emph{5}}Department of Physics \& Astronomy, University of Waterloo, Waterloo, ON, Canada\\
\textsuperscript{\emph{6}}Perimeter Institute for Theoretical Physics, Waterloo, ON, Canada
}

\date{\today}

\begin{abstract}
Echoed conditional displacement (ECD) gates for bosonic systems have become the
key element for real-time quantum error correction beyond the break-even point.
These gates are characterized by a single complex parameter $\beta$, and can be
constructed using Gaussian pulses and free evolutions with the help of an
ancillary transmon qubit. We show that there is a lower bound for the gate time
in the standard construction of an ECD gate. We present a method for optimizing
the pulse shape of an ECD gate using a pulse-shaping technique subject to a set
of experimental constraints. Our optimized pulse shapes remain symmetric, and
can be applied to a range of target values of $\beta$ by tuning only the
amplitude. We demonstrate that the total gate time of an ECD gate for a small value of
$\beta$ can be reduced either by relaxing the no-overlap constraint on the
primitives used in the standard construction or via our optimal-control
method. We show a slight advantage of the optimal-control method by
demonstrating a reduction in the preparation time of a $\ket{+Z_\mathrm{GKP}}$ logical state by
$\thicksim$$10\%$.
\end{abstract}

\maketitle

\section{ Introduction } \label{sec:1}
Bosonic codes encode logical qubits in the infinite Hilbert space of bosonic
modes represented by harmonic oscillators~\cite
{10.1103.revmodphys.77.513}. Several encoding schemes have been proposed, such
as cat codes~\cite{Bergmann.2016, Li.2017}, binomial codes~\cite
{Michael.2016}, rotation-symmetric codes~\cite{Grimsmo.2020}, and
Gottesman--Kitaev--Preskill (GKP) codes~\cite{Gottesman.2001, Fluhmann.2019,
Campagne-Ibarcq.2020}, the last of which have recently attracted a lot of
interest. GKP codes are designed such that they mitigate errors that cause small
displacements in the position or momentum of the quantum state in a
harmonic oscillator~\cite{royer2022}. The physical qubit encoded in a storage
cavity can be stabilized by protocols such as measurement-based feedback
loops~\cite{Campagne-Ibarcq.2020, Neeve.2022} or the ``small-Big-small''
(sBs) and ``Big-small-Big'' (BsB) protocols~\cite{Royer.2020}. These
stabilization protocols have been shown to increase the coherence time of the encoded
logical qubits~\cite{Sivak.2023, lachance2024autonomous}. It is anticipated
that this layer of bosonic quantum error correction at the physical level,
concatenated with conventional quantum error-correcting codes such as the
surface code~\cite{Fowler.2012bkc, Andersen.2020, AI.2023}, will reduce
the hardware overhead for fault tolerance~\cite{Noh.2020}.

Achieving universal control of bosonic systems is challenging and requires the bosonic
system to be coupled to a nonlinear component such as an atom~\cite
{law.1996}, a SNAIL~\cite{Grimm.2020}, or a qubit~\cite{Gao.2019}. One possible
primitive nonlinear operation resulting from an oscillator--ancilla system
operated in the dispersive regime is an echoed conditional displacement
(ECD) gate, which together with ancilla rotations form a universal gate
set~\cite{Campagne-Ibarcq.2020, Eickbusch.2021}. The preparation of quantum
states is then realized by optimizing parameterized quantum circuits composed
of alternating ancilla rotations and ECD gates~\cite{Hastrup.2021g2,
Sivak.2023}. Such a preparation protocol typically ignores the dependency of the
ECD gate time to its defining complex parameter $\beta$. The circuit optimization treats ECD
gates with different values of $\beta$ as though they are of equal weight.
Shortening an ECD gate based on the conditional displacement amplitude
$|\beta|$ has been proposed~\cite{Sivak.2023}, although not achieved for
small values of $\beta$. Nevertheless, the problem of optimizing the ECD gate
times while respecting experimental constraints has not been systematically studied.

In this paper, we present the results of our study of the relation between the
gate time and the parameter $\beta$ defining an ECD gate applied on a
three-dimensional cavity dispersively coupled to an ancillary transmon qubit
while considering experimental constraints for primitive pulses for qubit
rotations and cavity displacements. Given these experimental constraints, we
present an analytic lower bound and an approximate gate time while taking into account experimental constraints for the gate time as a function of the target
$\beta$ for the standard construction~\cite{Eickbusch.2021} of an ECD gate. We
then numerically optimize ECD gate pulses, first using the standard
construction, second by relaxing the no-overlap constraint applied to the
primitives, and finally with optimal-control pulse-shaping techniques, and compare
these pulses to the predicted values~\cite{Sivak.2023} of the gate time.
The predicted values can be reached by a standard ECD gate in the regime
$\beta >0.55$ and for all values of $\beta$ by both the ECD gate with
overlapping primitives and the pulse-shaping ECD gate we develop. We then
demonstrate the use of the studied relation between the target $\beta$ and
optimal gate time by optimizing a preparation protocol for a GKP logical state
faster than state-of-the-art protocols.

This paper is organized as follows. In Section~\ref{sec:2}, we present an
analysis of the time of an ECD gate given a target $\beta$ and a set of
experimental limits, and provide an analytic approximation. This approximate
gate time can be reached for the standard ECD gate~\cite{Eickbusch.2021} by using
numerical simulations. Then, the shortest ECD gate time is computed while
considering first-order nonlinearity and photon loss. Each gate is optimized
with respect to a set of parameters used to construct a standard ECD gate.
We then relax the no-overlap constraint applied to the Gaussian primitives
constructing the standard ECD gate to reduce the gate time in the low $\beta$
regime. In Section~\ref{sec:3}, we use a quantum optimal-control technique to
optimize an ECD gate and show that shorter ECD gates are attainable for small
values of $\beta$ compared to the standard ECD gate. In Section~\ref{sec:4}, we present the result of our
concatenation of the optimized ECD gates to prepare a logical GKP state, and compare
it with prior art~\cite{Eickbusch.2021}.

\section{\label{sec:2} Analysis of a Standard ECD Gate}

In this section, we derive an approximate gate time for an ECD gate based
on the analytic lower bound provided in Ref.~\cite{Eickbusch.2021}. In the standard
construction method~\cite{Eickbusch.2021}, each ECD gate is composed of four
displacement gates on the cavity, a $\pi$ rotation of the qubit, and two
free evolutions of the system resulting in conditional rotations. Typically,
the displacement gates and the $\pi$ rotation are each realized by a Gaussian
pulse. We compute the approximate gate time of an ECD gate constructed from
Gaussian pulses by considering a set of experimental constraints. For
simplicity, we ignore the effects of photon loss and higher-order
nonlinearities in the system Hamiltonian.

\subsection{\label{sec:2.1} The Approximate Gate Time of a Standard ECD Gate}

In the rotating frame at the drive and qubit frequencies, the time-dependant
Hamiltonian with the cavity's driving pulse is
\begin{equation}
    H(t)/\hbar=\Delta \hat{a}^\dagger_\mathrm{c}\hat{a}_\mathrm{c} - \chi \hat{a}^\dagger_\mathrm{c}\hat{a}_\mathrm{c} \hat{a}^\dagger_\mathrm{q}\hat{a}_\mathrm{q} + \varepsilon^*(t)\hat{a}_\mathrm{c} + \varepsilon(t)\hat{a}^\dagger_\mathrm{c},
\end{equation}
where $\hat{a}^\dagger_{\mathrm{c/q}}$ and $\hat{a}_{\mathrm{c/q}}$ are
the creation and annihilation operators, respectively, acting on the cavity and
the qubit, $\chi$ is the dispersive coupling strength, $\Delta = \chi/2$, and
$\varepsilon(t)$ is a complex-valued pulse shape acting on the cavity. The
time-dependent cavity's driving pulse is described by a pulse envelope $\varepsilon(t)$, which multiplies a carrier frequency of \mbox{$\omega =
(\omega_\mathrm{c}^\mathrm{g} + \omega_\mathrm{c}^\mathrm{e})/2$,} where
$\omega_\mathrm{c}^\mathrm{g}$ and $\omega_\mathrm{c}^\mathrm{e}$ are the
ground state's energy of the cavity with the qubit in the ground and excited
states, respectively, denoted by $\ket{g_\mathrm{q}}$ and $\ket{e_\mathrm{q}}$.

When projecting onto the first two levels of the qubit, the Hamiltonian becomes
\begin{equation}
    H(t)/\hbar = \chi \hat{a}^\dagger_\mathrm{c}\hat{a}_\mathrm{c} \hat{\sigma}_z/2 +\varepsilon^*(t)\hat{a}_\mathrm{c} + \varepsilon(t)\hat{a}^\dagger_\mathrm{c},
\end{equation}
where $\hat{\sigma}_z = \ketbra{e_{\mathrm{q}}} - \ketbra{g_\mathrm{q}}$. The action of
the $\pi$ pulse flipping the direction of rotation of the cavity's state in the
phase space in the middle of an ECD gate is represented by a function $z
(t)$~\cite{Eickbusch.2021}, which multiplies $\hat{\sigma}_z$, effectively
changing its sign. The modified Hamiltonian is then
\begin{equation}
    H(t)/\hbar = \chi \hat{a}^\dagger_\mathrm{c}\hat{a}_\mathrm{c} z(t)\hat{\sigma}_z/2 +\varepsilon^*(t)\hat{a}_\mathrm{c} + \varepsilon(t)\hat{a}^\dagger_\mathrm{c}.
\end{equation}
In the limit of instantaneous drives, $ z(t) = \pm 1$.

Following the mathematical derivation of Eickbusch et al.~\cite[Sec.~S4]{Eickbusch.2021}, we take the overall unitary
\begin{equation}
    U=\hat{\sigma}_xe^{i\theta'\hat{\sigma}_z/2}D(\lambda)C\!D(\beta)
\end{equation}
as an ansatz for the solution of the Schr\"odinger equation involving a
displacement gate $D(\lambda)$ and a conditional displacement gate $C\!D(\beta)$.

In order to realize the idealized target ECD gate, a displacement gate $D
(-\lambda)$, merged with the last displacement pulse constructing a standard
ECD gate, compensates for the spurious displacement gate with a parameter $\lambda$. Furthermore, a following virtual rotation
gate $R_z(-\theta')$ is applied to the resulting state to compensate for the geometric phase $\theta'$~\cite
{berry.1984, Eickbusch.2021}.

Referring still to the limit of instantaneous drives, the expression for the
final state's separation is then given by
\begin{equation}{\label{eq:inst_TECD_gate_time}}
    \beta=2i\alpha(t) \mathrm{sin}\hspace{-1mm}\left(\frac{\chi t}{2}\right)\approx i\alpha_0\chi t,
\end{equation}
where $\alpha_0=|\alpha(t)|_{\mathrm{max}}$ is the maximum displacement permitted on the cavity.

However, taking into account the finite bandwidth of the Gaussian pulses used as
primitives for the construction of the ECD gate leads to the more complete
representation
\begin{equation}
    z(t)=2\int_0^t \Omega(\tau) d\tau - 1,
\end{equation}
where
\begin{equation}
    \Omega(t)=\frac{1}{\sigma_\uppi\sqrt{2\pi}}\mathrm{exp}\hspace{-1mm}\left[-\frac{(t-T_\uppi/2)^2}{2\sigma_\uppi^2}\right]
\end{equation}
represents the Gaussian $\pi$ pulse applied to the qubit and $T_\uppi$ is the duration
of the $\pi$ pulse. The ECD gate time can then be approximated by
\begin{equation}{\label{eq:TECD}}
    T_{\mathrm{ECD}}\approx \frac{|\beta|}{\chi \alpha_0}+4\sqrt{2\pi}\sigma_\mathrm{c}+\sqrt{2\pi}\sigma_{\uppi}-4t_{\mathrm{tail}},
\end{equation}
where $\sigma_\mathrm{c}$ and $\sigma_{\uppi}$ are the standard deviation of the
Gaussian pulses constructing $\varepsilon(t)$ and $\Omega(t)$, respectively,
and $t_{\mathrm{tail}}=t_D/2-\sqrt{2\mathrm{ln}2}\sigma_\mathrm{c}$, where $t_D$ is the time needed to execute a displacement gate.
The experimental characterization of state brightening~\cite
{Eickbusch.2021} determines the value of $\alpha_0$. Using a value for
$\alpha_0$ that is too large will induce unwanted Kerr nonlinearity in the
cavity~\cite{Eickbusch.2021, lescanne.2019, sank.2016, reed.2010} and affect
the resulting gate fidelity. After deciding on a value for $\alpha_0$, we
compute $\sigma_\mathrm{c}$ given the maximum pulse amplitude. If we consider
the Lindblad superoperator $\mathcal{D}[\hat{a}_\mathrm{c}]$ for one of the
decoherence channels of the cavity, with a decoherence rate $\kappa_\mathrm
{c}$, then the displacement of the cavity's state by a microwave drive with a pulse envelope
$\varepsilon(t)$ is given by
\begin{equation}
    \alpha'(t)=-\gamma\alpha(t)-i\varepsilon(t),
\end{equation}
where $\gamma=i\frac{\chi}{2}+\frac{\kappa_\mathrm{c}}{2}$. Solving the above
equation yields
\begin{equation}{\label{eq:alpha_with_time}}
        \alpha(t)=-ie^{-\gamma t}\int_0^te^{\gamma \tau} \varepsilon(\tau)d\tau.
\end{equation}
We choose $\varepsilon(t)$ to be the Gaussian-shaped pulse
\begin{eqnarray}
    \mbox{$\varepsilon(t)=\varepsilon_0e^{-\frac{(t-t_D/2)^2}{2\sigma_c^2}}$},
\end{eqnarray}
where the total gate time is cut off at
$t_D = 2m_{\mathrm{cut}}\sigma_\mathrm{c}$, with $m_{\mathrm{cut}}$ being a
chosen parameter for the cut-off range. Next, we compute $\sigma_\mathrm{c}$ by
expanding Eq.~(\ref{eq:alpha_with_time}) to the first order and solving the
equation
\begin{equation}{\label{eq:determine_sigma}}
        \left|\frac{\gamma^2}{2}\sigma_\mathrm{c}^3-m_{\mathrm{cut}}\gamma \sigma_\mathrm{c}^2 + \sigma_\mathrm{c}\right| = \frac{\alpha_0}{\varepsilon_0\sqrt{2\pi}}.
\end{equation}
Substituting Eq.~(\ref{eq:determine_sigma}) into Eq.~(\ref{eq:TECD}), we obtain the
expression of the gate time for a standard ECD gate, taking experimental
constraints into account. The approximate gate time is obtained by fixing
$\alpha_0$ at the maximum value allowed by the experiment.

Choosing $\varepsilon_0=200$~MHz and $\alpha_0=30$, which are values used in
state-of-the-art experiments~\cite{Eickbusch.2021}, the Gaussian pulse for the
cavity is then calculated to be $\sigma = 11$~ns, resulting in a total length
of \mbox{$4\sigma = 44$~ns.} Figure~\ref{fig:beta_vs_T_infid}, in Section~\ref
{sec:3.3}, shows a comparison between the lower bound of \cite{Eickbusch.2021},
our approximate gate time obtained above, and the numerical results detailed in
Sections~\ref{sec:2.2}, \ref{sec:2.4}, and~\ref{sec:3.3}.
Figure~\ref{fig:beta_vs_T_infid} shows a constant gap independent of the
parameter $\beta$ between the lower bound of a standard ECD gate and
the approximate time we derive. This gap is explained by the ramp-up time of
the Gaussian pulses used as primitives for the cavity's displacements from the
experimental constraints of $\varepsilon_0$.

\subsection{The System Hamiltonian}

For all our optimization protocols, we consider a more complete representation
of the system Hamiltonian, which includes nonlinearities and the second-order
coupling term. This time-dependent Hamiltonian can be written as
\begin{eqnarray}
&&H(t)/\hbar\nonumber\\
&&=\Delta\hat{a}_\mathrm{c}^\dagger\hat{a}_\mathrm{c}-K_\mathrm{c}\hat{a}_\mathrm{c}^{\dagger2}\hat{a}_\mathrm{c}^2-K_\mathrm{q}\hat{a}_\mathrm{q}^{\dagger2}\hat{a}_\mathrm{q}^2\nonumber\\
&&~~~-\chi\hat{a}_\mathrm{c}^\dagger\hat{a}_\mathrm{c}\hat{a}_\mathrm{q}^\dagger\hat{a}_\mathrm{q}
-\chi'\hat{a}_\mathrm{c}^{\dagger2}\hat{a}_\mathrm{c}^2\hat{a}_\mathrm{q}^\dagger\hat{a}_\mathrm{q}\nonumber
+[\epsilon(t)\hat{a}_\mathrm{c}^\dagger+\epsilon^*(t)\hat{a}_\mathrm{c}]\\
&&~~~+[\Omega(t)\hat{a}_\mathrm{q}^\dagger+\Omega^*(t)\hat{a}_\mathrm{q}],
\end{eqnarray}
where $K_\mathrm{c}$ and $K_\mathrm{q}$ are the anharmonicity coefficients of
the cavity and qubit, respectively, and $\chi'$ is the second-order coupling
strength. Here, $H(t)$ acts on a Hilbert space $\mathcal{H} = \mathbb{C}^
{D_\mathrm{c}}\otimes\mathbb{C}^{D_\mathrm{q}}$, where $D_\mathrm{c}$ is the
truncated dimension of the Hilbert space of the cavity and $D_\mathrm{q}$ is
the truncated dimension of the Hilbert space of the qubit. The Hamiltonian
depends on a set of time-dependent driving fields $\varepsilon(t)$ and $\Omega
(t)$. The physical parameters chosen are from Ref.~\cite{Eickbusch.2021}.

\subsection{\label{sec:2.2} Optimizing a Standard ECD Gate}

We construct a standard ECD gate using truncated Gaussian pulses with a total
length of $4\sigma = 44$~ns as primitives for the cavity's displacement pulses
and one with a total length of $4\sigma = 24$~ns with a DRAG
term~\cite{Motzoi.2009} to perform the qubit's rotations.

For a chosen target $\beta$ and total gate duration $T$, the numerical
optimization of the ECD gate is performed in three steps and takes into
consideration the initial state \mbox{$\ket{\Psi} = \ket{0}_\mathrm
{c} \otimes \ket{+}_\mathrm{q}$} at each step of the process. First, the four
cavity displacement pulses are set to be equal in amplitude and the pulse amplitude that
minimizes the objective function
\begin{eqnarray}
    &&D^2(T/2) + D^2(T) + \alpha_{\mathrm{over}}^2(t=44 \mathrm{~ns}) \nonumber\\
    &&+~ \alpha_{\mathrm{over}}^2(t = T-44 \mathrm{~ns}) + (\beta - \beta_{\mathrm{target}})^2
\end{eqnarray}
in a closed system is found, where \mbox{$\alpha_{\mathrm{over}} =
|\alpha| - \alpha_{0}$} if $|\alpha| > \alpha_{0}$, and $0$ otherwise. This
objective function is constructed using identical terms to the ones in Ref.~\cite
{Eickbusch.2021}, with the exception that each individual term is squared to
alleviate the non-convexities of the objective function.

\begin{figure}[h!]
        \includegraphics[width=\linewidth]{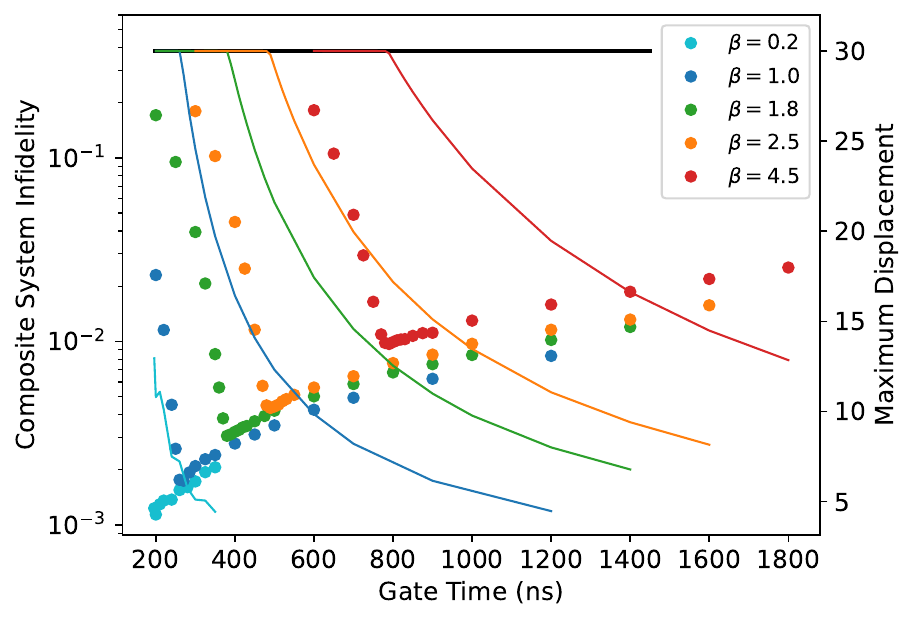}
        \caption{(left axis, dots) Best achieved infidelities for numerically optimized standard ECD gates with fixed target $\beta$ and gate durations. (right axis, curves) Maximum displacement reached for the optimized standard ECD gates' pulse sequence. For any chosen target $\beta$, an optimal gate duration is found at the smallest gate duration where the maximum displacement is not saturated by $\alpha_0$, shown using a black line. }
        \label{fig:OptimalInfidelity}
\end{figure}

A second round of optimization is then performed to minimize the infidelity of
the cavity in closed system using the objective function $(1-f_\mathrm
{c})^2 + \alpha_{\mathrm{over}}^2$, where $f_\mathrm{c}$ is the fidelity of the
cavity's state, and with the four cavity pulses fixed to an equal
amplitude. The third term in this objective function serves as a comparison of the maximum value
attained for $|\alpha|$ during the pulse sequence and is used to comply with the
experimental constraint $\alpha_0$ when the gate is too short to achieve
the target $\beta$. For a gate duration greater than or equal to the
optimal value, this term converges to zero during optimization
and, therefore, the optimization simply minimizes the infidelity. Finally, the
amplitude of each of the four cavity displacement pulses is treated as a
variational parameter to minimize the infidelity of the composite system
considering an open system and using the objective function
\begin{equation}
    (1-f)^2 + \alpha_{\mathrm{over}}^2
    \label{eq:cost_func_composite}
\end{equation}
and the Broyden--Fletcher--Goldfarb--Shanno (BFGS)
algorithm~\cite{fletcher.2013}. The qubit's geometric phase is then corrected
by applying a virtual rotation around the z-axis on the resulting state.
In Fig.~\ref{fig:OptimalInfidelity}, the infidelity resulting from our standard
ECD gate's optimization is shown as a function of gate time for a selection of
values of the target $\beta$. For any chosen target $\beta$, an optimal gate
duration can be found at the smallest gate duration where the maximum
displacement achieved during the gate is not saturated by $\alpha_0$. For gate
durations shorter than the optimal gate duration, the increasing infidelities
are due to the saturation of $\alpha$, which causes a mismatch between the
achieved $\beta$ and the target $\beta$. For gate durations longer than the
optimal gate duration, the increasing infidelities are due to the increasing
impact of the decoherence channels for longer ECD gate durations.

For every target $\beta$, the optimal gate duration is found and the optimized
pulse's infidelity is computed (see Fig.~\ref{fig:beta_vs_T_infid}; we remind
the reader that this plot appears in a later section because it is compared
with results of numerical optimizations yet to be described). This infidelity
is computed by averaging the gate infidelity applied over 96 initial
quantum states that form an overcomplete basis of displaced coherent states.
For $\beta < 0.55$, the gate duration is fixed at $T=196$~ns, which is the
minimum gate duration possible given the considered pulse primitives. The
amplitude of the cavity's displacement pulses is then reduced to achieve the
target $\beta$. Attempting to reduce the gate duration below our fixed minimum
of $T=196$~ns would result in Gaussian pulses overlapping with each other and
the qubit's $\pi$ pulse, which can skew the overall pulse shape. This prevents
the standard ECD gate from reaching the approximate gate time derived in
Section~\ref{sec:2.1} in the regime of $\beta < 0.55$
(see Fig.~\ref{fig:beta_vs_T_infid}). For $\beta \geq 0.55$, the gate duration
is set to the smallest possible value that complies with the $\alpha_0$
constraint, which, as shown in Fig.~\ref{fig:OptimalInfidelity}, yields the best gate infidelity in our numerical
simulations. In that regime, our
optimized standard ECD pulses closely follow the approximate gate times
(see Fig.~\ref{fig:beta_vs_T_infid}).

\begin{figure}[h]
        \includegraphics[width=\linewidth]{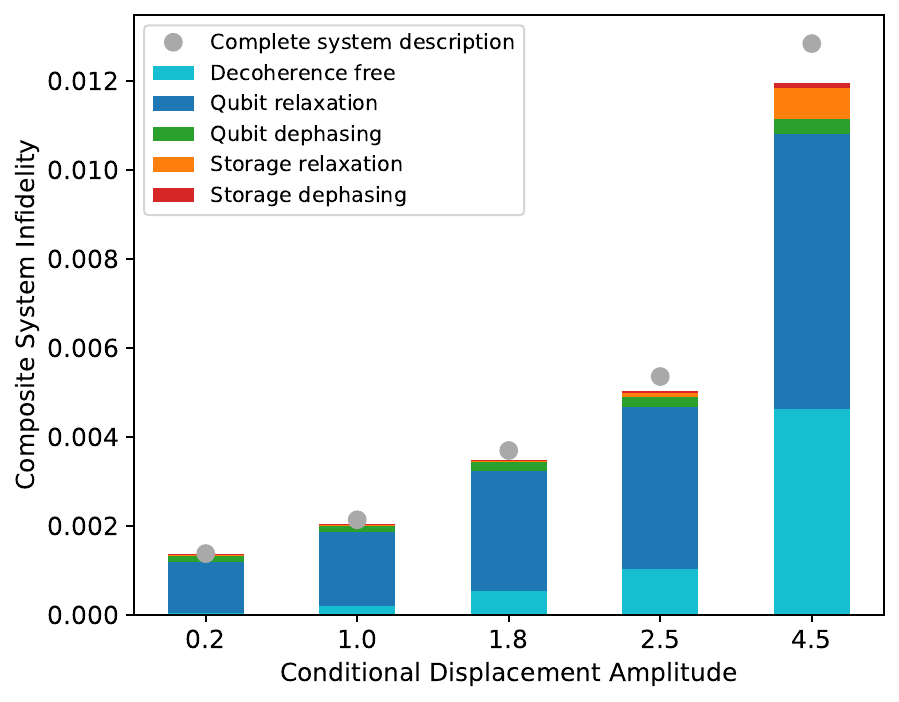}
        \caption{Error budget: the effect of decoherence channels on the composite system's infidelity for a single ECD gate for selected values of the target displacement amplitude. The infidelity of each decoherence channel is obtained by including a single decoherence channel at a time in each of the simulations and removing the decoherence-free contribution.}
        \label{fig:ErrorBudget}
\end{figure}

Figure~\ref{fig:ErrorBudget} shows the error budget of our standard ECD gate's
construction, computed by including a single decoherence channel at a time in
each of the simulations and removing the decoherence-free contribution from the
infidelity. The infidelity is dominated by the qubit's relaxation rate for
all values of $|\beta|$, but an increasing effect of the cavity's decoherence
channels and decoherence-free mechanisms is observed with increasing value of $|\beta|$. A
small mismatch is also observed between the sum of each decoherence channel's
contribution and the total infidelity of all the decoherence channels. This
could be explained by the compounding effects of having certain combinations of
decoherence channels.

\subsection{\label{sec:2.4} Relaxing the No-Overlap Constraint}

We now propose an alternative construction for an ECD gate, where primitives are
allowed to overlap with one another. By relaxing the no-overlap constraint that
was applied to our standard ECD gate's construction, the pulse can be optimized
to a shorter gate time in the $\beta < 0.55$ regime, as shown in Fig.~\ref{fig:beta_vs_T_infid}. We call these pulses ``overlapping primitives pulses'' (OP pulses).

For this optimization, the starting time for the second and third Gaussian
primitives applied to the cavity, $t_2$ and $t_3$, are added to the variational
parameters for the optimization. The first and last primitives remain locked at
the start and at the end of the ECD gate, respectively. The parameters are
optimized against the objective function Eq.~(\ref
{eq:cost_func_composite}), which is used in the final optimization round of
the construction of our standard ECD gate. Given a target $\beta$, we iteratively optimize the
parameters while reducing the total gate time by steps of $\Delta t = 10$ ns,
until no gain is observed in the gate fidelity. The resulting parameters in each
round of optimization are used as initial conditions for optimizing the
gate given the subsequent target gate time, with the exception of $t_3$, for which the
initial value is reduced by $\Delta t$ at every step. The $\pi$ pulse remains
locked at the centre of the gate.

Figure~\ref{fig:beta_vs_T_infid} shows the gate time as a function of target conditional displacement amplitude and the corresponding pulses' infidelities. The curve no longer
plateaus in the small $|\beta|$ regime and instead follows the approximate
gate times closely for all values of $|\beta|$. The second and third Gaussian primitives
now overlap with the $\pi$ pulse applied on the qubit, but no significant
impact is observed in the infidelity for all values of the conditional displacement amplitude considered.

\section{Optimal Control for an ECD Gate}
\label{sec:3}

In this section, we present an optimal-control approach that reduces the problem
of constructing ECD gates to an optimization problem. We then compare the
resulting pulse shapes to the OP pulse shapes. In addition, we compare the
resulting gate times to the approximate and standard gate times as described
above.

\subsection{The Optimization Problem}

Due to the truncated dimensions of the cavity's Hilbert space, not all quantum
states will evolve properly under the given truncated Hamiltonian, for example,
the state $\ket n$ for $n \gg 1$. Therefore, we choose the average infidelity over a
subset $\mathcal S$ of operator basis states as our objective function. We choose this
approach over more-standard objective functions, such as the average
gate fidelity over a complete basis set~\cite{Nielsen.2002}, which are
computationally too costly, as they require a large number of evaluation rounds of the objective function. An alternative objective function could
be the infidelity between a target unitary operator and the resulting unitary
operator. However, the computed unitary operator may again be unreliable due to
the truncated dimensions of the cavity, and is not easily parallelizable.

To determine $\mathcal S$, we first construct a pulse shape for a standard ECD
gate for the target $\beta$. Then, the product states of the overcomplete
coherent states for a bosonic system~\cite{Klauder.1960} and the operator basis
states for a qubit~\cite{Nielsen.2002} are chosen as the initial states. After
evolving the initial states using the pulse shape of a standard ECD gate, we
select the initial states that result in an infidelity that is lower than a selected
threshold to be members of $\mathcal S$. Thus, the goal of the optimization
problem is to find the optimal fields that drive the ensemble of initial states
$\{\ket{\Psi_\mathrm{init}^s}\}_{s \in \mathcal S}$ into the corresponding
target states $\{\ket{\Psi_\mathrm{target}^s} \}_{s \in \mathcal S}$ in a total
gate time $T$, with an objective function $\mathcal F$ that penalizes
deviations from the following constraints: the maximum displacement shift
$|\alpha|_\mathrm{max}$ in the cavity to not be greater than a threshold
$\alpha_0$ and the pulse shape vanishing at the both the start and the end of the
gate. We define our objective function as
\begin{equation}
    \mathcal{F}_T(\vec{c}) = \sum_{i=1}^3 f_{i, T} (\vec c),
\end{equation}
where $T$ is the fixed total gate time, $\vec{c}$ represents all the
coefficients constructing the pulse shape in Eq.~(\ref{eq:20}) in Subsection~(\ref{sec:3.2}), and the
objective functions $f_i$ are defined as follows.

\begin{itemize}
\item  The first objective function is the average infidelity over
$\mathcal S$:
\begin{equation}
    f_{1,T}(\vec c)= \frac{1}{|\mathcal S|}\sum_{s\in \mathcal S}
    \left[1-\left|\braket{\Psi^s_{\mathrm{target}}}{\Psi^s_T(\vec c)}\right|^2\right].
    \label{eq:f1}
\end{equation}
\item The maximum displacement shift of the cavity must be smaller than the
 threshold $\alpha_0$, so we use a sigmoid function to penalize for the values
 of $|\alpha|$ that exceed the threshold, which leads to the objective
 function
\begin{equation}
    f_{2, T}(\vec c)= C_1/[1+\mathrm{exp}(|\alpha|_{\mathrm{max}}(\vec{c})-\alpha_0)],
    \label{eq:f2}
\end{equation}
where $C_1$ is a hyperparameter.

\item The displacement shift of the cavity at the end of the evolution is
penalized so as to stay close to zero so that in consecutive applications of
ECD gates (e.g., in preparing GKP states as described in Section~\ref{sec:4}) the value
$|\alpha|$ remains small at both the beginning and the end of each ECD gate:
\begin{eqnarray}
\hspace{-1em} f_{3, T}(\vec c) =& C_2 |\alpha(T, \vec{c})|^2 + C_3 |\varepsilon(0, \vec{c})|^2 + C_4 |\varepsilon(T, \vec{c})|^2.
\label{eq:f4}
\end{eqnarray}
\end{itemize}
Note that the control 
amplitude of $\varepsilon(0)$ and of  $\varepsilon(T)$ are included in the cost 
function to force them to be as close to zero as possible following the transformations applied during 
the pulse shape construction. Also note that we optimize pulse shapes for 
$\alpha(0) = 0$.

\subsection{Construction of Pulse Shapes}
\label{sec:3.2}

For the sake of computational efficiency, control pulses can be parameterized by a set of
basis functions, for example, Fourier functions or B-spline functions. Following Ref.~\cite{Petersson.2020}, we use quadratic B-spline basis functions with equidistant
centres as the basis functions such that no more than three of them are nonzero
at any point in time. This allows the control functions to be evaluated
efficiently. The parameterized pulse shapes can be written as $\varepsilon
(t) = \varepsilon^x(t) + i\varepsilon^y(t)$, with
\begin{equation}
{\varepsilon^x(t)}=\sum_{m=1}^Dc_m^xB_m(t) \quad \mathrm{and} \quad
{\varepsilon^y(t)}=\sum_{m=1}^Dc_m^yB_m(t),
\label{eq:20}
\end{equation}
where $c_m^x$ and $c_m^y$ are real coefficients, $B_m(t)$ are the B-spline
functions, and $D$ is the number of B-spline functions and will be determined later. Furthermore, the
maximum amplitude of the pulses is bounded by applying the transformation
$\mathrm{tanh}(\varepsilon_x(t))$ to the original pulses.

The pulses applied on the cavity are also are chosen to be symmetric around the
midpoint $t= T/2$. Therefore, the pulse shape at the time segment $[0, T/2]$
is mirrored and a butterfly low-pass filter of order 6 with a bandwidth set to
$\Delta f = 50$~MHz is applied to the generated pulse. This achieves two
outcomes: the resulting pulse shape is smooth at the midpoint and the
bandwidth of the pulses abides by experimental limitations.

For each target $\beta$, we set the initial gate time to be equal to the time of
an optimized standard ECD gate. To determine the number of basis functions $D$,
we fit the above-mentioned ansatz to the standard ECD gate pulse shape and
choose the smallest possible value for $D$ that matches the infidelity of the standard
ECD gate. We do this to reduce the experimental overhead and decrease the total computation time of the optimization process.

Because the ECD gate is only defined up to a virtual $Z$ gate, the control 
pulse of the qubit is fixed to be a single $\pi$ pulse as is done in the standard
construction. Optimizing the entire $\Omega(t)$ pulse results in pulse shapes without any relevant contribution to the resulting unitary evolution (not shown in this paper). 

Because we have a larger number of variables for the optimization problem
compared to the standard construction, we use the limited-memory variant of
BFGS with boundaries (\mbox{L-BFGS-B})~\cite{liu.1989} implemented in
SciPy~\cite{virtanen2020scipy}, and we pass auto-differentiated gradients
of the objective function to the optimizer using Jax~\cite{jax2018git}.
For each round of optimization given a target $\beta$, the pulse shape is
randomly initialized. We then use the resulting parameters in the subsequent
optimization round, but for a shorter pulse shape with a new total time
$T-\Delta t$, using $\Delta t = 10$~ns. For a sufficiently small value of
$\Delta t$, the optimization for a shorter gate time nearly always converges
faster than an optimization with a random initialization of the parameters.

\subsection{Numerical Results}
\label{sec:3.3}

\begin{figure}[h]
        \includegraphics[width=\linewidth]{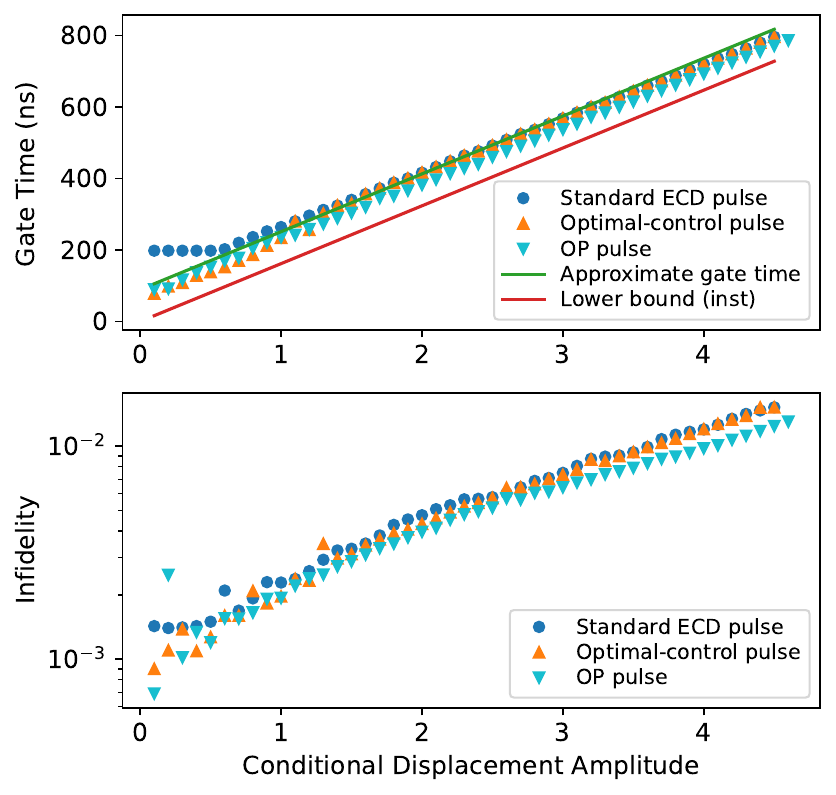}
        \caption{Comparison of optimized pulse shapes for ECD gates constructed using the standard method~\cite{Eickbusch.2021} (indicated by blue dots), with overlapping primitives allowed (cyan  triangles) and using a pulse-shaping method (orange triangles). (top) Comparison of gate time between the numerically optimized pulses and the approximate gate time computed with experimental constraints (green line) and the lower bound with instantaneous drives (red line, the data for which is obtained from Eq.~(\ref{eq:inst_TECD_gate_time}) and identical to that of Ref.~\cite{Eickbusch.2021}). (bottom) Comparison of the infidelity in open system between the numerically optimized standard ECD pulses, the OP pulses, and optimal-control pulses.}
        \label{fig:beta_vs_T_infid}
\end{figure}

The gate time and infidelity as a function of the parameter $\beta$ resulting
from our pulse-shaping method are shown in Fig.~\ref{fig:beta_vs_T_infid}. In
the regime $\beta \geq 1$, our optimal-control pulses, the standard pulses, and
the OP pulses converge to having almost identical gate times.
The small discrepancy between the OP pulses and the optimal-control
pulses is explained by the fact that the optimal-control parameters are initialized
randomly for the first choice of $T$ given each $\beta$, while the OP pulses are initialized from a physically well-informed guess.
However, in the regime $\beta < 1$, the optimal-control pulses outperform the
other pulse shapes by a small amount and outperform the approximate gate time derived
in Section~\ref{sec:2.1}. In particular, in the regime $\beta \leq 0.55$, the
optimal-control pulses continue to improve the gate time as $\beta$ is decreased without compromising the infidelity of any of the resulting gates.

\begin{figure*}[ht!]
        \includegraphics[width=\linewidth]
        {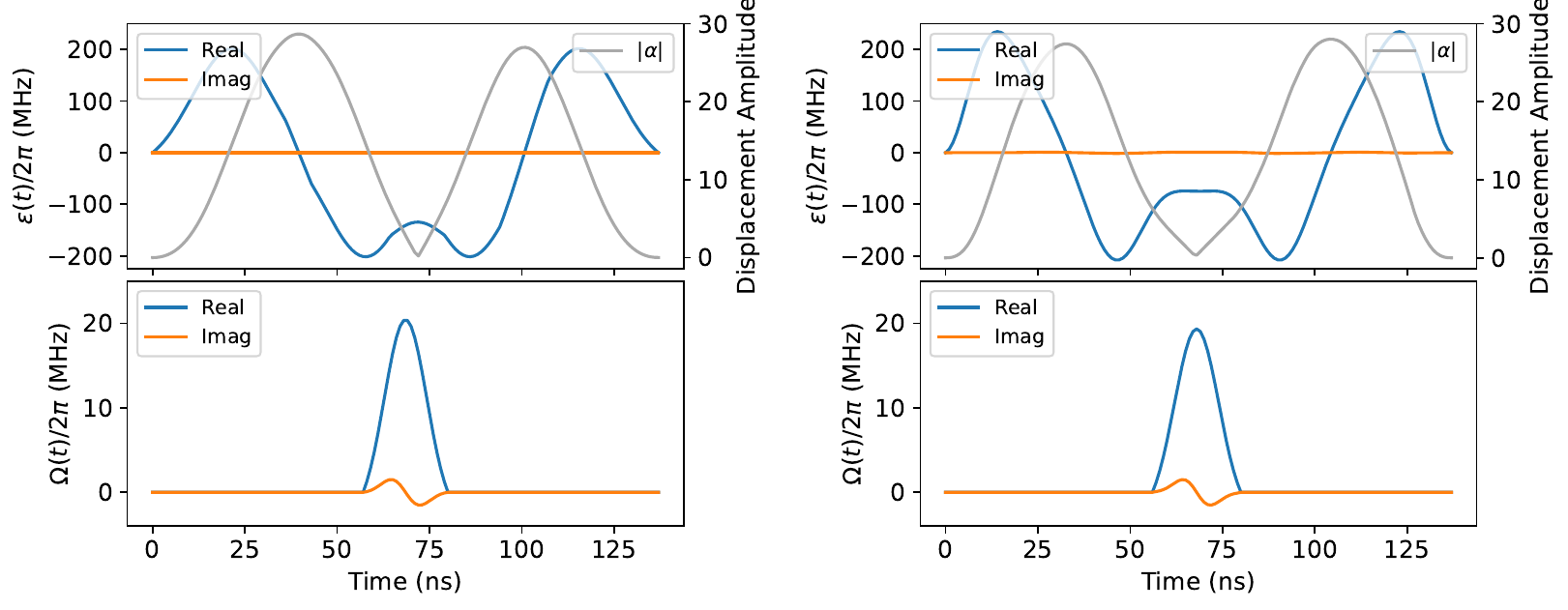}
        \caption{Pulse shapes for $\varepsilon(t)$ and $\Omega(t)$ resulting from (left) the overlapping primitives' construction described in Sec.~\ref{sec:2.4} and (right) the pulse shaping method described in Sec.~\ref{sec:3.2}. The pulses were optimized for $\beta = 0.4$ and with the constraints $\alpha_0 = 30$ and $\varepsilon_0 = 200$~MHz. The displacement amplitude $|\alpha|$ resulting from the cavity's pulse $\varepsilon(t)$ is shown in grey (right axis).}
        \label{fig:beta_0d4}
\end{figure*}

Figure~\ref{fig:beta_0d4} (right side) shows the pulse shape optimized for the ECD gate with
a target $\beta=0.4$. We observe that, similar to the case of OP pulse shapes (Fig.~\ref{fig:beta_0d4}, left side), the two Gaussian pulses in the vicinity of the
middle point and the $\pi$ pulse on the qubit overlap in time. This combination
reduces the total gate time. Note that the qubit's $\pi$ pulse is applied while
the cavity is displaced from the vacuum state. However, in the standard
construction, we avoid applying a $\pi$ pulse when the cavity is displaced, as
it may be miscalibrated due to the displacement of the cavity and the dispersive
coupling. However, given the fact that the final average state's fidelity of
both the ECD gate with overlapping primitives and the pulse-shaping ECD gate
are better than that of the standard ECD gate, we conclude that this effect does
not obstruct the performance of our pulses. The overlap between different
primitive-like pulses in the optimal-control pulses explains the small offset
between our gate times and the approximate gate times derived in Sec.~\ref
{sec:2.1}.

In addition, the phase space trajectories generated by our optimal-control
pulses are similar to those of both the standard and OP pulses, as shown by the resulting displacement amplitude
$|\alpha(t)|$ throughout the evolution (see the grey curve in Fig.~\ref
{fig:beta_0d4}). The pulse shapes consist of a fast displacement far from the
origin, a free evolution if needed, and another displacement to return the
state to the origin in the phase space before applying the $\pi$ pulse on the
qubit and repeating another sequence of two displacements. These trajectories, along with the results
shown in Fig.~\ref{fig:beta_vs_T_infid}, suggest that the ECD gate with
overlapping primitives is a ``bang--bang'' control scheme~\cite{rolewicz.2013}.

\section{\label{sec:4}Circuit Optimization for the Preparation of a GKP State}
In this section, we showcase how the optimized pulse shapes for ECD gates with
small values of $\beta$ can be utilized to reduce the total preparation time
of a $\ket{+Z_{\mathrm{GKP}}}$ logical state. This state
has been reported to be prepared in approximately $3.3~\upmu$s with a quantum circuit
composed of nine ECD gates and 10 single-qubit rotations \cite{Eickbusch.2021}. In
Fig.~\ref{fig:beta_vs_T_infid}, it is shown that the gate time is nearly
proportional to the magnitude of the parameter $\beta$ of the ECD gate. To
minimize the GKP state preparation time, we take advantage of these shorter
ECD gates by incentivizing the choice of smaller values of $\beta$ for a
sequence of $N= 9$ ECD gates using the objective function
\begin{equation}{\label{eq:prep_loss}}
    L = 1 - \mathrm{tr}(\hat{\rho}_\mathrm{c}\hat{\rho}_{\mathrm{c,target}})
    + C\sum_{i=1}^N{|\beta_i|}, 
\end{equation}
where $C$ is a hyperparameter. The reduced density matrix for
the cavity at the end of the state preparation is \mbox{$\hat{\rho}_\mathrm
{c} = \mathrm{tr}_\mathrm{q}\hat{\rho}_{\mathrm{cq}}$}, and the target cavity
state is \mbox{$\hat{\rho}_{\mathrm{c,target}} = \ketbra{+Z_{\mathrm{GKP}}}$.}

\begin{figure*}
        \includegraphics[width=\linewidth]
        {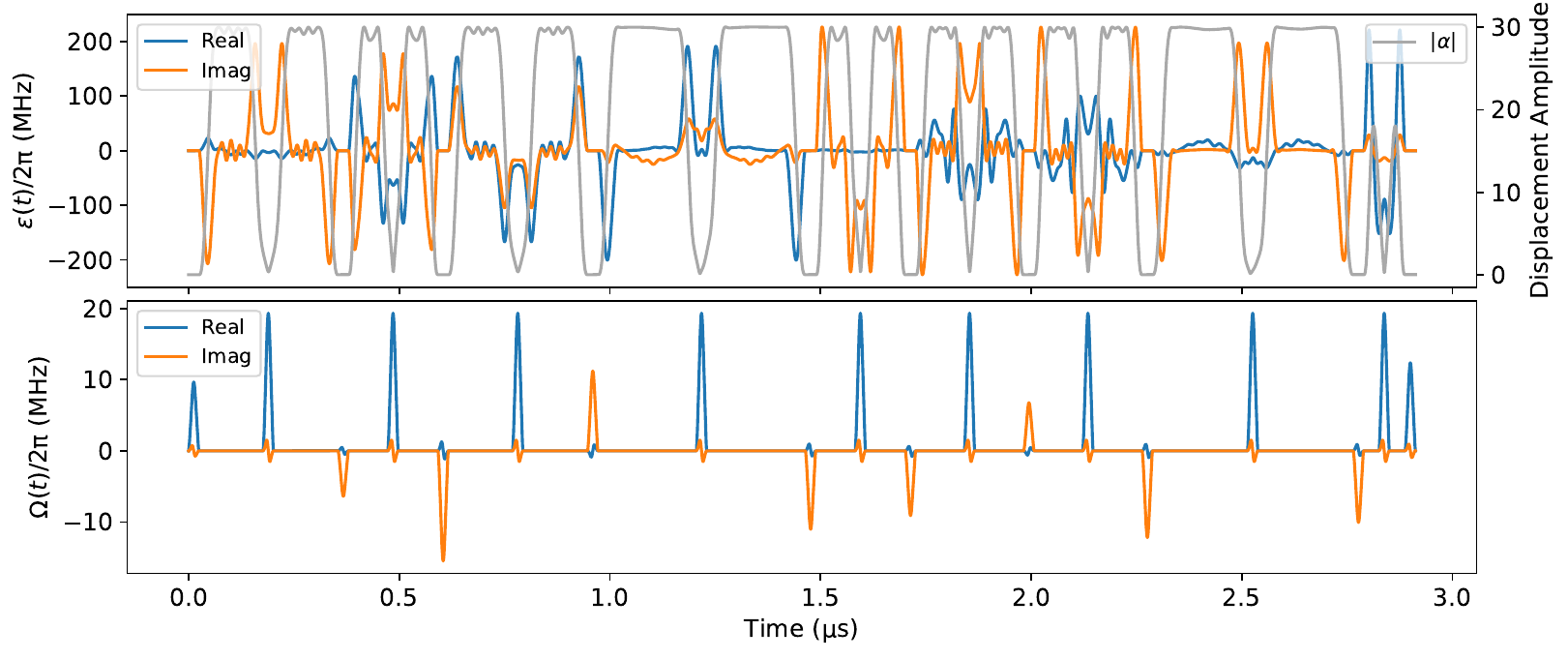}
        \caption{Optimized pulse sequence using a circuit depth of $N=9$ to prepare a $\ket{+Z_{\mathrm{GKP}}}$ logical state with the parameter \mbox{$\Delta_{\mathrm{target}}=0.306$~\cite{Royer.2020}} in the cavity, using a vacuum state as the initial state. The ECD gates' control parameters \{$\beta$, $\varphi$, $\theta$\} for this pulse sequence are shown in Table~\ref{table:controlparameters}.}
        \label{fig:pulse_sequence}
\end{figure*}

After obtaining the optimized circuit parameters for the preparation protocol, the
optimized ECD gates' pulses for the resulting values of $\beta$ are
concatenated to realize fast preparation of a GKP state. We construct each
ECD gate by multiplying the optimized pulse shape for each real value of
$\beta$ by a phase coefficient. A virtual rotation gate $R_z$ is applied after each ECD
gate, prior to each subsequent qubit rotation, to compensate for the geometric
phase, which is not taken into account in our pre-optimized ECD gates from
the previous section. The entire sequence for preparing a $\ket{+Z_{\mathrm
{GKP}}}$ state is shown in Fig.~\ref{fig:pulse_sequence} and the corresponding $\{
\beta$, $\varphi$, $\theta$\} parameters are shown in Table~\ref
{table:controlparameters}. The total preparation time is $2.912~\upmu$s and the
closed-system cavity fidelity is 0.978.

By applying the optimized pulse shapes for ECD gates, the preparation sequence
can be reduced by over 300~ns. Given that the gate time of ECD gates for 
large values of $\beta$ is close to that of  standard ECD gates, only the ECD gates with
small value of $\beta$ are replaced by  optimal-control pulses, in order to reduce
experimental characterization efforts.

\section{\label{sec:conc}Conclusion}

We have provided a detailed analysis of a standard ECD gate and compared it with
an alternative construction where primitives are allowed to overlap and with
optimal-control pulses that result from our numerical optimizations. Both the
OP pulses and the optimal-control pulses provide a small
improvement in the gate time of an ECD gate for small values of $\beta$, and
our results suggest that the current standard pulse is near optimal otherwise.
Using our optimized pulses for small values of $\beta$, a greater number of
stabilization rounds can be performed on GKP states, as the
state-of-the-art protocols, such as the sBs protocol~\cite{Royer.2020}, include
ECD gates with small values of $\beta$ which are executed many times during an
experiment.

\section*{ACKNOWLEDGEMENTS}

We thank our editor, Marko Bucyk, for his careful review and editing of the
manuscript. We thank Marc-Antoine Lemonde, Dany Lachance-Quirion, and Julien
Camirand Lemyre at Nord Quantique for helpful discussions on numerical
simulations of a bosonic system and the construction of a standard ECD pulse
shape. \mbox{M.~L.-M.} further acknowledges the financial support of Mitacs. P.~R. acknowledges the financial support of Mike and Ophelia
Lazaridis, Innovation, Science and Economic Development Canada (ISED), and the
Perimeter Institute for Theoretical Physics. Research at the Perimeter Institute
is supported in part by the Government of Canada through ISED and by the
Province of Ontario through the Ministry of Colleges and Universities.



\begin{table*}[h!]
\begin{center}
\begin{tabular}{|c|c|c|c|}
\hline
j  & $\varphi_\mathrm{j}/\pi$ & $\theta_\mathrm{j}/\pi$ & $\beta_\mathrm{j}$ \\ \hline
1  & $0.00$                  & $0.50$                    & $-0.13+1.54i$      \\ \hline
2  & $0.50$                  & $-0.33$                   & $-0.56+0.75i$      \\ \hline
3  & $-0.50$                 & $0.80$                    & $-1.26-0.81i$      \\ \hline
4  & $-0.50$                 & $-0.58$                   & $2.42+0.54i$       \\ \hline
5  & $-0.50$                 & $0.57$                    & $-0.01-0.93i$      \\ \hline
6  & $0.50$                  & $-0.47$                   & $-0.26+1.15i$      \\ \hline
7  & $-0.50$                 & $-0.35$                   & $0.33-1.13i$       \\ \hline
8  & $0.50$                  & $-0.63$                   & $-0.22+2.41i$      \\ \hline
9  & $-0.50$                 & $0.52$                    & $-0.23-0.03i$      \\ \hline
10 & $0.00$                  & $0.64$                    & $0$                  \\ \hline
\end{tabular}
\end{center}
        \caption{Best set of control parameters \{$\beta$, $\varphi$, $\theta$\} found for the preparation of a $\ket{+Z_{\mathrm{GKP}}}$ state with the parameter \mbox{$\Delta_{\mathrm{target}}=0.306$~\cite{Royer.2020}} in the cavity. This set of parameters is found by minimizing Eq.~(\ref{eq:prep_loss}) with randomized initial parameters more than $10^8$ times.}
\label{table:controlparameters}
\end{table*}

\clearpage
\bibliography{references}

\end{document}